# Carbon Nanotubes as Intracellular Protein Transporters:

# Generality and Biological Functionality

Nadine Wong Shi Kam and Hongjie Dai[*]

Department of Chemistry and Laboratory for Advanced Materials, Stanford University,

Stanford, CA 94305, USA

Email: hdai@stanford.edu

Various proteins adsorb spontaneously on the sidewalls of acid-oxidized single-walled carbon nanotubes. This simple non-specific binding scheme can be used to afford non-covalent protein-nanotube conjugates. The proteins are found to be readily transported inside various mammalian cells with nanotubes acting as the transporter via the endocytosis pathway. Once released from the endosomes, the internalized protein-nanotube conjugates can enter into the cytoplasm of cells and perform biological functions, evidenced by apoptosis induction by transported cytochrome $c$. Carbon nanotubes represent a new class of molecular transporters potentially useful for future in-vitro and in-vivo protein delivery applications.

## Introduction

The interaction between nanostructured materials and living systems is of fundamental and practical interest and will determine the biocompatibility, potential utilities and applications of novel nanomaterials in biological settings. The pursuit of new types of molecular transporters is an active area of research, due to the high impermeability of cell membrane to foreign substances and the need for intercellular delivery of molecules via cell-penetrating transporter for drug, gene or protein therapeutics.[1-3] Recently, we and others[4-8] have uncovered the ability of single-walled carbon nanotubes (SWNTs) to penetrate mammalian cells and further transport various cargos inside cells including small peptides,[5] the protein streptavidin[4] and nucleic acids.[6,8] In our work of nanotube internalization and streptavidin transporting using nanotube carriers[4] and the work of Cherukiri et al.[7] on nanotube uptake, the internalization mechanism was attributed to endocytosis. In the work of Pantarotto et al.[5], Bianco et al.[8] and Lu et al.,[6] nanotube uptake was suggested to be via insertion and diffusion through the lipid bilayer of cell membrane. While the uptake mechanism awaits a consensus, it has been consistently reported that well-processed water-soluble nanotubes exhibit no apparent cytotoxicity to all living cell lines investigated thus far, at least in a time frame of days.

Here, we report a finding that SWNTs are generic intracellular transporters for various types of proteins ($\leq$ 80 kD) non-covalently and non-specifically bound (NSB) to nanotube sidewalls. The proteins investigated include streptavidin (SA), protein A (SpA), bovine serum albumin (BSA) and cytochrome *c* (cyt-*c*). The intracellular protein transporting and uptake via nanotube carriers are also generic for various adherent and non-adherent mammalian cell lines including HeLa, NIH-3T3 fibroblast, HL60 and Jurkats cells. Energy dependent endocytosis is confirmed to be the internalization mechanism. Further, with cytochrome *c* as the cargo protein, we present an exploration of the fate of internalized protein-SWNT-protein conjugates, attempts of releasing the conjugates from the endosome vesicles into the cell cytoplasm using chloroquine, and investigation of the biological functions of the released proteins. We observe apoptosis or programmed cell death induced by cyt-*c* transported inside cells by SWNTs after release from the endosomes. The results provide the first proof of concept of in vitro biological functionality and activity of proteins delivered by SWNT molecular transporters.

## Materials and Methods

**Materials.** Alexa-fluor 488-labeled streptavidin (SA, molecular weight 60 kD), alexa-fluor bovine serum albumin (BSA, 66 kD) were purchased from Molecular Probes Inc. Protein A (42 kD), cytochrome *c* (12 kD) and Fluorescein isothiocyanate-labeled human immunoglobin G (hIgG, 150 kD) were obtained from Aldrich.

**Fluorescent labeling of proteins.** Cyt-*c* and SpA were fluorescently labeled with Alexa-fluor 488 moieties by a protein labeling kit obtained from Molecular Probes Inc. In brief, a protein solution at a concentration of 2 mg/mL in standard phosphate buffer saline was mixed with 50 μL of sodium bicarbonate solution and the provided vial of Alexa- fluor dye and reacted for 1 h at room temperature. After reaction, the protein-dye conjugate was flowed through a gel separation column (Bio-rad Biogel P-30) for purification. The resulting fluorescently labeled protein solution was then diluted to a concentration of ~ 10 μM in PBS.

**Purification, cutting and oxidation of SWNTs.** Similar to the steps described previously,[4,9,10] SWNTs (20 mg) grown by laser ablation were mixed with 100 mL of 2.5 M $HNO_3$, refluxed for about 36 h, sonicated with a cup-horn sonicator (Branson Sonifer 450) for 30 min to cut the nanotubes into short segments and refluxed again for another 36 h. After this treatment, the mixture was filtered through a polycarbonate filter (Whatman, pore size 100 nm), rinsed thoroughly and then re-suspended in pure water by sonication. The aqueous suspension was then centrifuged at 7000 rpm (revolutions per minute) for about 5 min to remove any large impurities from the solution. SWNTs after these processing steps were in the form of short (tens to hundreds nanometers) individual tubes (~1.5 nm in diameter) or small bundles (up to 5 nm in diameter) and re-suspended to give a concentration of ~0.04-0.05mg/mL. Acidic oxygen groups (e.g., -COOH) on the sidewalls of the tubes rendered solubility or high suspension stability of the SWNTs in water and buffer solutions.

**Conjugation of proteins to SWNTs.** A suspension of the oxidized and cut SWNTs at a concentration of ~ 0.05 mg/mL was mixed with fluorescently labeled proteins (typical protein concentration ~1μM) for ~ 2 h at room temperature prior to characterization (by e.g., atomic force microscopy (AFM) for imaging protein-SWNT conjugates) or cellular incubation for uptake. After this simple mixing step, proteins were found to adsorb non-specifically onto nanotube sidewalls.

**AFM characterization of protein-SWNT conjugates.** AFM was used to study non-specific binding of proteins on nanotubes. After SWNT-protein conjugation, a drop of the solution (~50 μL) was pipetted onto a clean $SiO_2$ substrate and allowed to stand for

~ 30-45 min. The substrate was then rinsed with copious amount of HPLC grade $H_2O$ and dried by a $N_2$ stream.

**Incubation of living cells in solutions of protein-SWNT conjugates.** Non-adherent HL60 and Jurkat cells were both grown in RPMI-1640 cell culture medium (Invitrogen) supplemented with 10% fetal bovine serum (FBS). Adherent HeLa and NIH-3T3 cells were grown in DMEM cell medium (Invitrogen) supplemented with 10% FBS and 1% penicillin-streptomycin. The cell density for all cell lines used in the incubation was ~ $3x10^6$ cells/mL.

For incubating HL60 and jurkat cells in nanotube solutions, 100 μL of the cell suspension was typically mixed with 100 μL of a protein-SWNT conjugate solution at 37 °C for 2-3 h in 5% $CO_2$ atmosphere. The concentration of SWNTs in the incubation solution was typically ~0.05 mg/mL. After incubation, the cells were washed, collected by centrifugation and re-suspended in the cell culture medium twice.

For incubating HeLa and NIH-3T3 cells in nanotube solutions, the cells were seeded into 8-well chambered cover slides ~24 h prior to incubation in a protein-SWNT solution in DMEM cell growth medium. The incubation conditions were the same as the one for non-adherent cells above. Since the cells were adhered to the cover-slide surface, no centrifugation was needed to separate the cells from the incubation solution. The cells were washed by changing the cell medium and then characterized by confocal microscopy imaging.

**Confocal microscopy imaging of cells after incubation in solutions of fluorescently labeled protein-SWNT conjugates.** All confocal images of cells were recorded with a Zeiss LSM 510 confocal microscope immediately (except for cell

viability assay described later) following the incubation in solutions of protein-SWNT conjugates and washing steps.  In the case of non-adherent cells, 20 µL of the cell suspension were pipetted onto a glass cover slide before imaging.  Adherent cells were imaged directly in chambered cover slides.

**Flow Cytometry.**  The cells were analyzed by a Becton-Dickinson FACScan instrument after incubation for 2-3 h in a solution of fluorescently labeled protein-SWNT conjugate.  The cell suspension was supplemented with 2% propidium iodide (PI, Fluka chemicals) to stain dead cells.  The data presented here represent the mean fluorescence obtained from a population of 10,000 cells.  Note that the mean fluorescence reported was a measure of fluorescence of live cells only as cells showing high levels of PI staining were excluded from the analysis.

**Cell proliferation MTS Assay.**  After incubation with the SWNT-streptavidin conjugate (unlabeled), HL-60 cells were washed and resuspended in RPMI media.  The cells were plated in a 6-well plate at a density of ~ $8 \times 10^3$ cells/well.  At 24 h interval, Celltiter96 reagent (Promega) was added to one of the wells and allowed to incubate for ~ 2-3 h at 37 ºC and 5% $CO_2$.  The CellTiter 96 assay uses the tetrazolium compound (3-(4,5-dimethylthiazol-2-yl)-5-(3-carboxymethoxyphenyl)-2-(4-sulfophenyl)-2H-tetrazolium, inner salt; MTS) and the electron coupling reagent, phenazine methosulfate (PMS). MTS is chemically reduced by cells into formazan, which is soluble in tissue culture medium. The measurement of the absorbance of the formazan can be carried out at 490nm. The assay measures dehydrogenase enzyme activity found in metabolically active cells.

**Apoptosis Assay.** We use Annexin V labeled with fluorescein isothiocyanate (Molecular Probes) as an early stage apoptosis marker. One of the earliest indications of apoptosis is the translocation of the membrane phospholipid phosphatidylserine (PS) from the inner to the outer leaflet of the plasma membrane. Once exposed to the extracellular environment, binding sites on PS become available for Annexin V, a ~35 kDa $Ca^{2+}$-dependent, phospholipid binding protein with a high affinity for PS[11] (also see Sigma Technical Bulletin number MB-390). In this study, NIH 3T3 cells were plated and incubated with SWNT-cytochrome *c* as described above. After incubation, the cells were washed, trypsinized to detach them from the plate surface and washed with phosphate buffer saline. Annexin V-FITC was added to the cell suspension in the presence of the binding buffer and allowed to react for 20 min at room temperature. The cells were co-stained with propidium iodide and immediately analyzed by flow cell cytometry. Apoptosis data in Fig. 6 were for cells free of PI-staining recorded ~ 4-5 h after exposure to chloroquine with or without (for control) cyt *c*-SWNT conjugates.

**Results and Discussion**

**Protein binding on oxidized SWNTs.** The combined treatment of refluxing and sonication in nitric acid is known to produce short (50-500nm, Fig. 1a) individual or small bundles of SWNTs with oxygen containing groups (e.g., -COOH) along the sidewalls and ends of the tubes.[4,9,10] These functional groups impart hydrophilicity to the nanotubes and make them stable in aqueous solutions (in pure water and various buffers including PBS and cell culture media) without apparent aggregation in the timescale monitored (1-3 h). In the current work, we found that simple mixing of the oxidized

SWNTs with protein solutions led to non-specific binding of proteins to the nanotubes as can be gleaned from the AFM data in Fig. 1b-d for BSA, SpA and cyt-*c*. AFM imaging revealed that the average spacing between proteins adsorbed on SWNTs ranged approximately from ~20-100 nm and the loading appeared to be the highest for cyt-*c* (Fig. 1d) likely due to attractive electrostatic interactions (isoelectric point pI ~ 9.2 for cyt-*c* and the oxidized SWNTs were negatively charged). Proteins with pI < 7 such as SA and BSA also exhibited affinity for oxidized SWNT sidewalls (Fig.1b, 1c). We attributed the binding to either electrostatic forces between functional groups on SWNTs and positively charged domains on proteins, and/or hydrophobic interactions since the SWNT sidewalls were not fully oxidized (~tens of nanometer between oxygen groups) and contained hydrophobic regions.

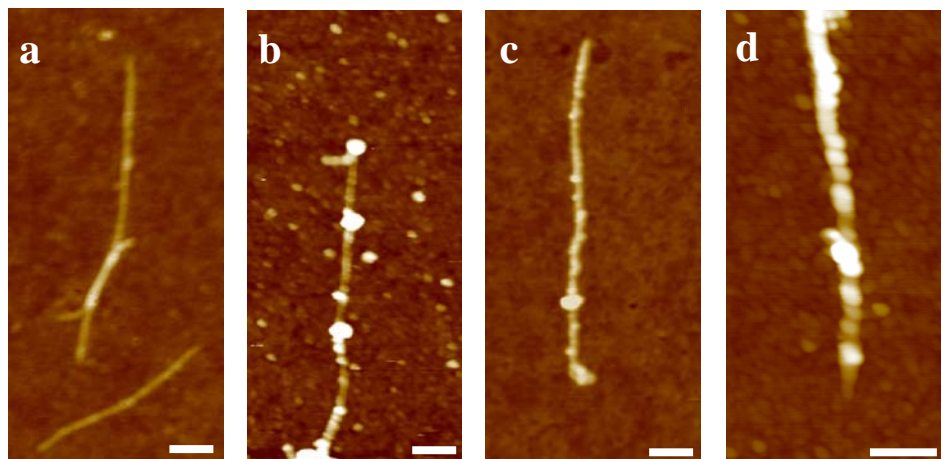

**Figure 1. AFM images of various SWNT samples deposited on SiO$_2$ substrate.** (a) Oxidized SWNT prior to conjugation with proteins and after conjugation to 1 μM of (b) Alexa-Fluor 488 BSA, (c) Alexa-Fluor 488 spA and (d) Alexa-Fluor 488 cytochrome C. Scale bar = 100 nm.

We and others previously observed a general phenomenon of protein non-specific binding on as-grown and acid-oxidized SWNTs mainly due to hydrophobic interactions.[12-15] In agreement with Green et al.,[14] we found that imparting hydrophilicity to SWNTs by oxidation was insufficient to block protein NSB. This result was utilized from here on to create non-covalent protein-SWNT conjugates for cellular uptake. Note that in our earlier protein/SWNT intracellular transport, binding between nanotube transporters and SA was obtained through biotin-streptavidin conjugation.[4]

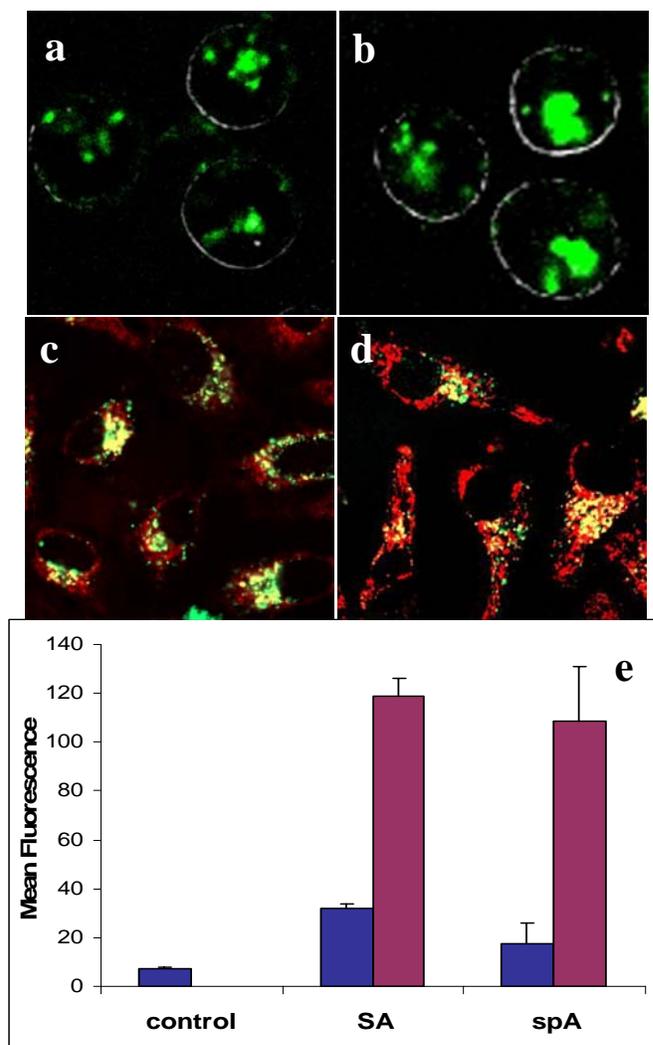

**Figure 2.** Confocal microscopy and flow cytometry characterization of cells after incubation in protein-SWNT (proteins labeled to be green fluorescent) solutions for 2 h. (a) HL60 cells after incubation in streptavidin SA-SWNTs (b) HL60 cells after incubation in BSA-SWNTs. (c) HeLa cells after incubation in SA-SWNTs in the presence of the FM 4-64, a red membrane and endocytotic vesicle marker. Yellow color in the image is due to co-localization of fluorescently labeled green proteins and red FM 4-64. (d) HeLa cells after incubation in cytochrome *c*-SWNTs in the presence of FM 4-64. Note that in Fig.2c&2d, the cell nucleus (dark circular or oval-shaped regions) appears free of fluorescence with the internalized protein-SWNT conjugates accumulating outside of the nucleus region and outlining parts of the nucleus boundary. (e) Cell cytometry data for untreated HL60 cells (labeled as 'control'), and HL60 cells incubated in solutions of fluorescently labeled SA and SpA respectively (blue bars) and treated with the respective fluorescent protein-SWNT conjugate (purple bars).

**Cellular uptake of protein-SWNT conjugates.** To investigate the fate of the SWNT-protein conjugates in-vitro by standard characterization tools, fluorescently labeled proteins by alexa-fluor 488 (excitation wavelength ~488 nm and emission wavelength ~ 510 nm) were used. The proteins were conjugated to SWNTs in concentration ranging from 100 nM - 1μM with SWNT concentration of ~ 0.05mg/mL. For various non-adherent cells (HL-60 and jurkat) and adherent cell lines (HeLa and NIH-3T3), we found that intracellular internalization of protein-SWNT conjugates was generic for the various proteins investigated by confocal microscopy imaging (Fig. 2a-2d) and flow cytometry (Fig. 2e). Optical confocal microscopy sections (in the Z-direction) of the cells revealed that fluorescence was mainly originated within the cell interior, though cell membrane-surface bound fluorescence of protein-SWNT conjugates was also present. As control experiments, we carried out incubations of cells in solutions that contained fluorescently labeled proteins alone and compared the detected fluorescence level with cells exposed to protein-SWNTs by flow cytometry. As shown in Fig. 2e, the fluorescence level detected for the former was low compared to the latter, suggesting that while proteins in solutions were unable to traverse across cell membranes by themselves, SWNTs were effective in transporting protein cargos inside cells.

The observed internalization of the non-covalently bound proteins via oxidized SWNT transporters was similar to that of streptavidin transported by biotinylated SWNTs for which endocytosis was identified as the internalization pathway.[4] Endocytosis is a well known mechanism for a wide range of species traversing cell membranes including large liposomes and nanoparticles, and is an energy-dependent internalization mechanism hindered at low temperatures.[16-18] Indeed, by incubating cells in protein-SWNT conjugates at 4 ºC, we observed no uptake of the conjugates (Fig. 3), suggesting the endocytosis mechanism for the cellular uptake of protein-nanotube conjugates in experiments performed at 37 ºC.

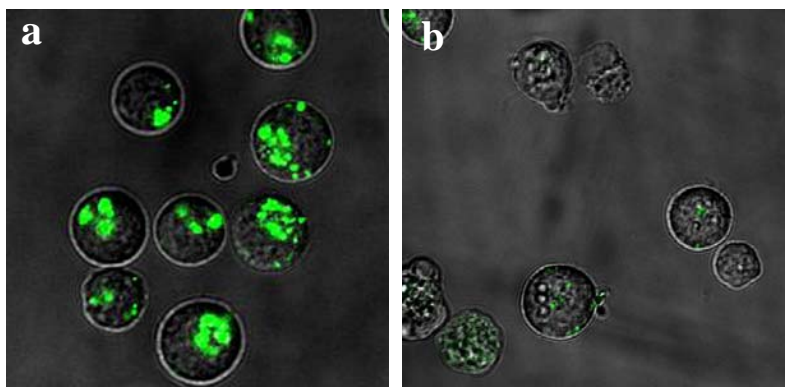

**Figure 3.** Experiments at various temperatures. Confocal images of HL60 cells after incubation in cytochrome *c*-SWNTs for 2 hrs at (a) 37 °C and (b) 4 °C respectively

We are currently investigating details of the endocytotic uptake of SWNT conjugates to address whether the conventional endocytosis mechanism involving clathrin coated pits[19] on cellular membranes are involved for the nanotube uptake. We speculate that SWNTs exhibit high binding affinity to certain cell membrane species that facilitate the high efficiency of nanotube binding and subsequent internalization. This binding affinity exists even after protein adsorption on SWNTs since the protein coverage is often incomplete[4] and the nanotube/protein complex still exhibit substantial hydrophobicity. The uptake mechanism certainly warrants much future effort and we will present our detailed mechanistic study in a separate publication. Discrepancy in uptake mechanisms for nanotubes, i.e., endocytosis found by us and phagocytosis proposed by Cherukiri et. al.,[7] versus insertion and diffusion by Pantarotto et al. and Bianco et al.[5,8] needs to be reconciled.

We note that while binding and intracellular protein transporting by SWNTs appeared general for small to medium sized proteins (molecular weight <= 80 kD), cellular uptake of protein-SWNT conjugates was poor and nearly non-existent for a large protein investigated, i.e., human immunoglobulin (molecular weight ~150kD). This observation (data not shown) is not understood currently and is likely related to the large size of the antibody cargo, which may have caused inefficient loading of hIgG on SWNTs due to the large size mismatch (~7-8nm for hIgG; ~1.5 nm in diameter for SWNT) or inefficient endocytosis due to the large conjugates. The issue of binding and transporting of large proteins and antibodies with SWNTs is open for further investigation.

**Cell proliferation after nanotube uptake by MTS assay.** Biocompatibility is a major concern when introducing any foreign substances inside living systems. Thus far, several groups have reported that relatively pure, well solubilized short carbon nanotubes appear nontoxic once internalized into mammalian cells without apparent adverse effects to cell viability.[4-8] In the current work, we further assessed the biocompatibility of carbon nanotubes by monitoring cell proliferation. We used the CellTiter 96 MTS assay (Promega) to examine the proliferation of HL60 cells following exposure to SWNT-protein conjugates. The CellTiter 96 assay uses the tetrazolium compound (3-(4,5-dimethylthiazol-2-yl)-5-(3-carboxymethoxyphenyl)-2-(4-sulfophenyl)-2H-tetrazolium, inner salt; MTS) and the electron coupling reagent, phenazine methosulfate (PMS). MTS is chemically reduced by cells into formazan, which is soluble in tissue culture medium. The absorbance of the formazan salt can be measured at 490 nm and correlates to the number of cells in the suspension. The absorbance from the cells incubated with the SWNT-protein conjugates was measured over a period of 5 days in parallel with control cells never exposed to any nanotubes. As shown in Fig. 4, the absorbance detected by a UV-VIS spectrophotometer was very similar for cells with and without exposure to SWNTs over the 5 day monitoring period. This result suggest that cell proliferation and cell viability was unaffected by the internalized carbon nanotubes.

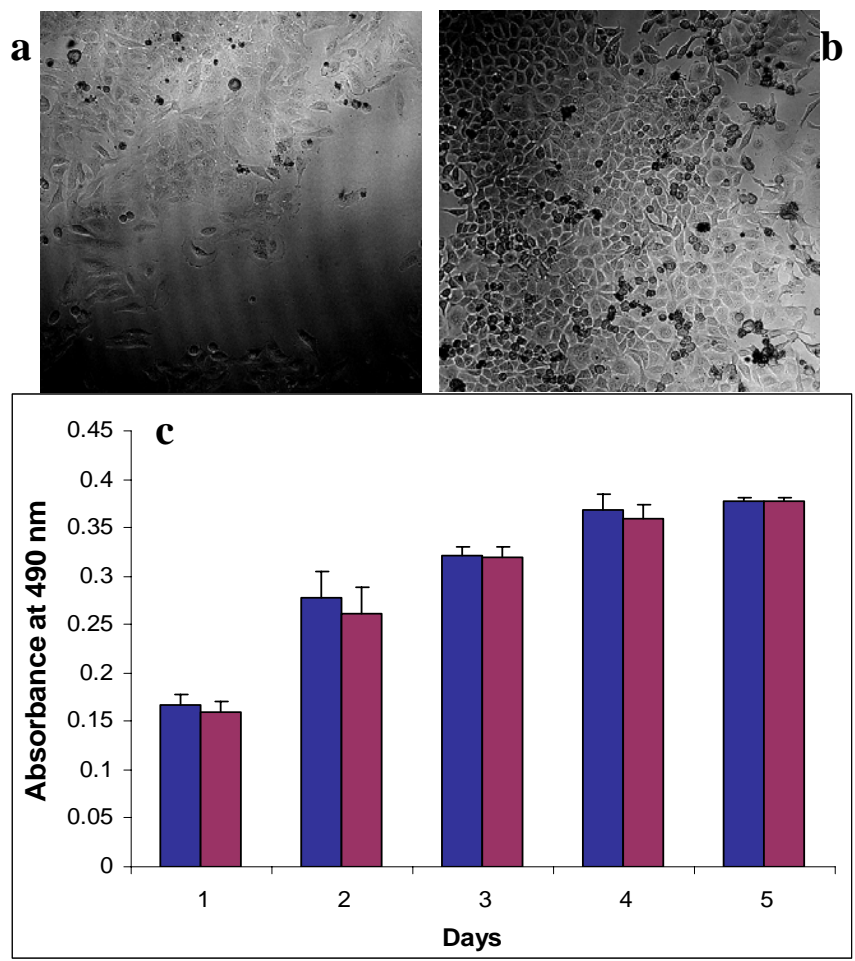

**Figure 4. Cell Proliferation Assay.** Cells are incubated with SWNT-SA conjugate (without fluorescent label) for 2 hrs. After incubation, cells are washed, collected and resuspended in cell media. Confocal images of HeLa cells taken (a) 24hrs and (b) 48hrs after incubation. The cells are plated at a cell density of 8 x $10^3$ cells/well and returned to a $CO_2$ incubator at 37 ºC and 5% $CO_2$. (c) At 24 hr interval after the initial SWNT incubation, CellTiter96 reagent is added to the cell suspension, and allowed to incubate for 2 hrs at 37 ºC. The absorbance is then taken at 490 nm. The proliferation of the cells incubated with SWNT is monitored for a period of up to 5 days after initial exposure to SWNT (red) and does not show any deviation from the proliferation of untreated cells (blue).

**Endosome release.** As observed previously, SWNT-protein conjugates, once internalized inside mammalian cells were colocalized with a red endocytosis endosome marker FM 4-64 (Fig. 2c, 2d),[4] corresponding to the containment of internalized species in endosomal lipid vesicle compartments (specifically labeled by red FM-64). Consistent with this was the observed punctuate fluorescence (Fig.2a-b) inside cells due to endosomal confinement of the internalized species. The endocytosis pathway can be illustratively described as the engulfment of the cell membrane and formation of a lipid vesicle around the species to be internalized.[16-18] Once inside the cells, the endosomes could fuse with the cell lysosomes, which may lead to later degradation of the internalized species in the lysosomes. To avoid the fate of lysosome degradation, it is important to trigger endosomal release of the internalized molecules into the cell cytoplasm. This will then open up the possibility of obtaining biological functionality for the internalized cargo molecules. Note that green fluorescence (proteins) non-overlapping with FM 4-64 markers were sometimes observed (Fig.2c-d), suggesting certain internalized species might have been released from the endosome via an intentional mechanism.

Various approaches have been suggested in the past to actively initiate endosomal release of endocytosed species including complexation of the cargo molecules with pH sensitive polymers or highly amine-rich moieties.[20,21] These methods make use of the pH difference across the endosome membrane and the pH response of the polymers or amines. For SWNT transporters, similar complexation schemes can be developed to afford nanotube transporters containing an endosome-releasing agent. While such an effort warrants detailed investigations separately from the current work, we describe here the release of internalized protein-SWNT conjugates from the endosomal compartments by adding chloroquine to cell medium during incubation of cells in protein-SWNT conjugates. Chloroquine is a membrane permeable base that has been shown to localize inside endosomes and cause an increase in pH.[22] The resulting osmotic pressure can lead to swelling of the endosomal compartments and eventual rupture.[23] In experiments carried out with and without chloroquine, we noticed a difference in the distribution of the detected fluorescence inside the cells. In the absence of chloroquine, discrete punctate fluorescence spots were observed within the cells (Fig. 5a), similar to those in Fig. 2 and 3. In contrast, when cells were simultaneously exposed to protein-SWNT conjugates and 100 μM chloroquine, fluorescence signals inside the cells appeared to be diffuse and uniform (Fig. 5b), due to redistributed conjugates over the entire cell cytoplasm after endosome releasing.

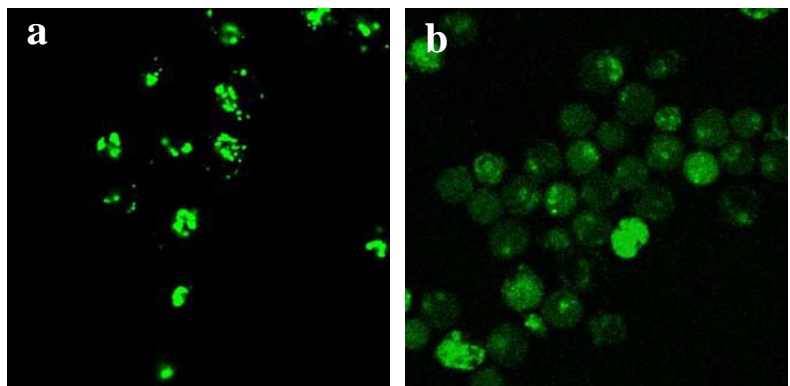

**Figure 5. Endosomal Rupture.** (a) Cells are incubated with (a) SWNT-CytC conjugate and (b) SWNT-CytC + 100 μM of chloroquine at 37 ºC and 5% $CO_2$. Confocal images are taken immediately after incubation and washing. Confocal images indicating the release of the SWNT-protein conjugate from the endosome, overall green color across the cell in (b) vs. green individualized spots inside the cells in (a).

**Towards biological functionality: apoptosis induction by SWNT transported cytochrome *c*.** Building upon the results above on binding and intracellular transporting of proteins with SWNT carriers and cargo releasing into cell cytoplasm, we carried out an investigation of biological functions of the transported proteins. We chose cyt-*c* for this exploration due to the high degree of loading of cyc-*c* on SWNTs (Fig.1d) and high efficiency of endocytosis of the cyt-*c*/SWNT conjugates (Fig.5a). Cyt-*c* is the smallest protein in the current work with a molecular weight of 12 kD and has the highest pI of ~9.2 to afford a high degree of binding onto the negatively charged oxidized SWNTs. It has been suggested that cells during apoptosis or programmed cell death may release cytochrome *c* from mitochondria.[24,25] When microinjected, cytochrome *c* has been shown to induce or activate apoptosis by bypassing the need for release of cytochrome *c* from mitochondria.[24] Through interaction with apoptotic protease activating factors (Apaf), cyt-*c* in the cytosol is believed to initiate the activation cascade of caspases that leads to cell death.[25]

HeLa and NIH-3T3 cell lines, shown previously to undergo cyt-*c*-induced apoptosis were investigated in this study for intracellular transporting of cyt-*c* with SWNTs and for apoptosis assay.[25] Unlike the proteins used for other uptake experiments, cyt-*c* used for binding to SWNTs and transported inside cells for apoptosis assay were free of any fluorescence label. After incubation of cells in the cyt-*c*-SWNT conjugates, apoptosis was analyzed using fluorescently (FITC) labeled Annexin V. Annexin V-FITC is an efficient marker for early stage apoptosis as it binds specifically to the exposed phospholipd phosphatidylserine (PS) translocated from the inner to the outer leaflet of the plasma membrane during apoptosis. Once exposed to the extracellular environment, binding sites on PS become available for Annexin V, a ~35 kDa $Ca^{2+}$-dependent, phospholipid binding protein with a high affinity for PS.[11]

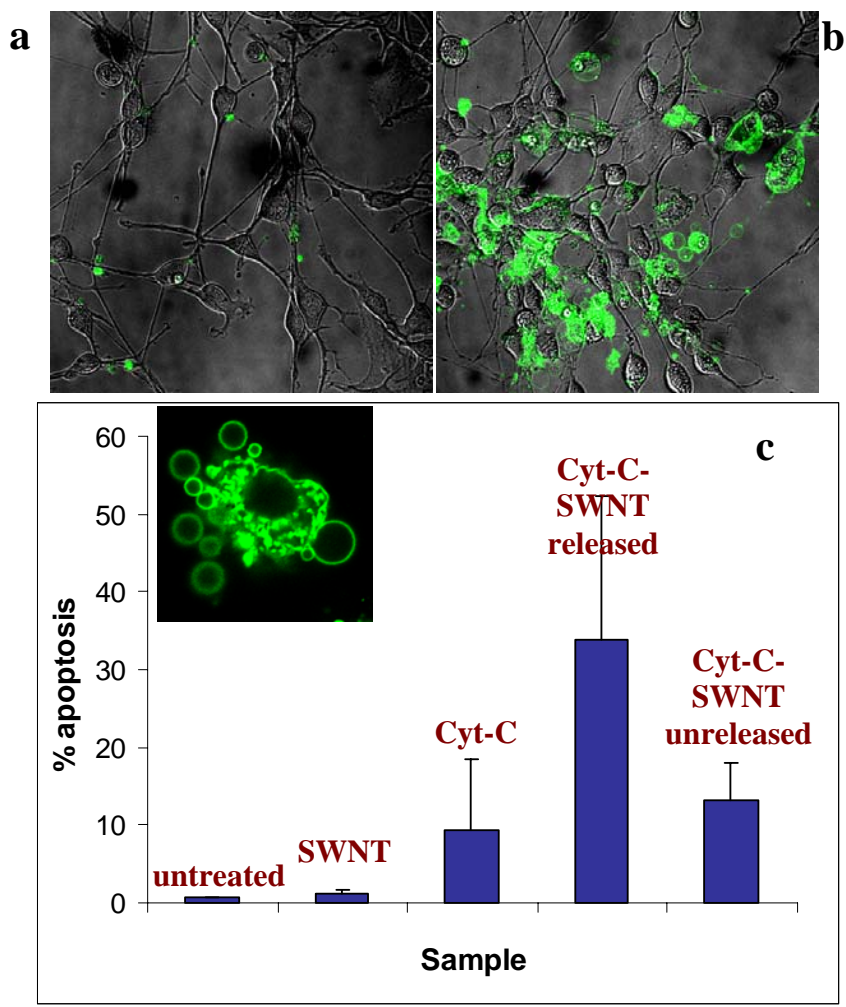

**Figure 6.** Apoptosis induction by cytochrome *c* cargos transported inside cells by SWNTs. (a) Confocal image of NIH-3T3 cells after 3 h incubation in 50 μM of cytochrome *c* alone (no SWNT present) and 20 min staining by Annexin V- FITC (green fluorescent). (b) Images of cells after incubation in 50 μM cytochrome c-SWNTs in the presence of 100 μM chloroquine and after Annexin V- FITC staining. (c) Cell cytometry data of the percentages of cells undergoing early stage apoptosis (as stained by Annexin V-FITC) after exposure to 100 μM of chloroquine only (labeled 'untreated'), SWNT +100 μM of chloroquine, 10 μM of cyt-*c* + 100 μM chloroquine, 10 μM of cyt-*c*-SWNT + 100 μM chloroquine and cyt *c*-SWNT without chloroquine. The inset shows a representative confocal image of the blebbing of the cellular membrane (stained by Annexin V-FITC) as the cell undergoes apoptosis. Note that PI co-staining was used and all data shown here excluded PI-positive cells and are recorded ~ 4-5 h after exposure to chloroquine. The level of PI staining for all cells here was a normal ~ 4-6% out of ~10.000.

For NIH-3T3 cells incubated with cyt-c alone and cyt-*c*-SWNT conjugates, we analyzed the degree of Annexin V staining by both confocal microscopy (Fig. 6a,6b) and cell flow cytometry (Fig. 6c). We observed significantly higher percentages of apoptosed cells incubated with cyt-*c*-SWNT conjugates than those incubated with cyt-*c* alone (Fig. 6a vs. 6b; also see Fig.6c). We also encountered apoptotic cells exhibiting blebbing of the cellular membrane stained by Annexin V-FITC (Fig.6c inset), a known phenomenon for cells undergoing apoptosis.

To investigate the effect of endosomal release on the efficiency of apoptosis induction by cyt-*c* transported by SWNTs, we carried out incubations of cells in cyt-*c*-SWNT with and without the presence of chloroquine. Higher degrees of apoptosis were consistently observed for cells treated with cyt-*c*-SWNT in the presence of chloroquine (Fig.6c) as attributed to the more efficient endosomal releasing of proteins. Note that the apoptosis results presented here were reproduced at least three times in independent experimental runs. Also, unless otherwise stated, chloroquine was used in all incubations including control experiments for fair comparisons with the positive control (cyt-*c*-SWNT + chloroquine incubation). Taken together, our data suggested that the cytochrome *c* bound and transported inside cells by SWNT carriers remained biologically active for apoptosis induction. It was unclear however, whether the cytochrome *c* functionality was obtained after detaching from the SWNT sidewalls or with the proteins bound to nanotubes. Future work is required to elucidate this issue.

**Conclusion**

We have found that various proteins adsorb spontaneously on the sidewalls of acid-oxidized SWNTs and this non-specific binding afford non-covalent protein-nanotube conjugates. Nanotubes appear to be generic intracellular protein transporters for various proteins and mammalian cell lines. Cellular uptake is via the energy dependent endocytosis pathway and the endocysed species are confined inside endosomes lipid compartments. Once released into the cytoplasm of cells, the proteins can perform biological functions as evidenced by apoptosis induction by the transported cytochrome *c*. The door is currently open for addressing outstanding issues and capturing opportunities presented by nanotube molecular transporters. The detailed uptake mechanisms should be established without ambiguity. Efficient cargo releasing strategies should be developed for nanotubes since the use of chloroquine here is only to illustrate the need of endosomal release and does not represent a practical approach to for real applications. Long term cytotoxicity effects should be established for nanotubes in vitro and in vivo. The uniqueness of SWNTs as molecular transporters is beginning to emerge and need to be fully developed. Carbon nanotubes could become a new class of molecular transporters for various in-vitro and in-vivo delivery applications.

**Acknowledgments.** We would like to thank Dr. Frederic B. Kraemer for kindly providing the NIH-3T3 cells. This work is supported by NSF Stanford CPIMA.

**References**

(1) Henry, C. M. *Chem. Eng. News* **2004**, *82*, 37.

(2) Smith, D. A.; vandeWaterbeemd, H. *Curr. Opion. Chem. Biol.* **1999**, *3*, 373.

(3) Bendas, G. *Biodrugs* **2001**, *15*, 215.

(4) Kam, N. W. S.; Jessop, T. C.; Wender, P. A.; Dai, H. J. *J. Am. Chem. Soc.* **2004**, *126*, 6850.

(5) Pantarotto, D.; Briand, J.; Prato, M.; Bianco, A. *Chem. Comm.* **2004**, *1*, 16.

(6) Lu, Q.; Moore, J. M.; Huang, G.; Mount, A. S.; Rao, A. M.; Larcom, L. L.; Ke, P. C. *Nano Lett.* **2004**, *4*, 2473

(7) Cherukuri, P.; Bachilo, S. M.; Litovsky, S. H.; Weisman, R. B. *J. Am. Chem. Soc.* **2004**, *126*, 15638.

(8) Bianco, A.; Hoebeke, J.; Godefroy, S.; Chaloin, O.; Pantarotto, D.; Briand, J.-P.; Muller, S.; Prato, M.; Partidos, C. D.; 16-Dec-2004, W. R. D. *J. Am. Chem. Soc.* **2005**, *127,* 58.

(9) Chen, J.; Hammon, M. A.; Hu, H.; Chen, Y. S.; Rao, A. M.; Eklund, P. C.; Haddon, R. C. *Science* **1998**, *282*, 95.

(10) Liu, J.; Rinzler, A. G.; Dai, H.; Hafner, J. H.; Bradley, R. K.; Boul, P. J.; Lu, A.; Iverson, T.; Shelimov, K.; Huffman, C. B.; Rodriguez-Macias, F.; Shon, Y.-S.; Lee, T. R.; Colbert, D. T.; Smalley, R. E. *Science* **1998**, *280*, 1253.

(11) Pigault, C.; Follenius-Wund, A.; Schmutz, M.; Freyssinet, J.; Brisson, A. *J. Mol. Biol.* **1994**, *236*, 199.

(12) Balvavoine, F.; Schultz, P.; Richard, C.; Mallouh, V.; Ebbeson, T. W.; Mioskowski, C. *Angew. Chem. Int. Ed.* **1999**, *38*, 1912.

(13) Shim, M.; Kam, N. W. S.; Chen, R.; Li, Y.; Dai, H. *Nano Lett.* **2002**, *2*, 285.

(14) Azamian, B. R.; Davis, J. J.; Coleman, K. S.; Bagshaw, C. B.; Green, M. L. H. *J. Am. Chem. Soc.* **2002**, *124*, 12664.

(15) Chen, R. J.; Bangsaruntip, S.; Drouvalakis, K. A.; Kam, N. W. S.; Shim, M.; Li, Y. M.; Kim, W.; Utz, P. J.; Dai, H. J. *Proc. Nat. Acad. Sci. USA.* **2003**, *100*, 4984.

(16) Silverstein, S. C.; Steinman, R. M.; Cohn, Z. A. *Annu. Rev. Biochem.* **1977**, *46*, 669.

(17) Vida, T. A.; Emr, S. D. *J. Cell Bio.* **1995**, *128*, 779.

(18) Mukherjee, S.; Ghosh, R. N.; Maxfield, F. R. *Physiol. Rev.* **1997**, *77*, 759.

(19) Schmid, S. L. *Annu. Rev. Biochem.* **1997**, *66*, 511.

(20) Lackey, C. A.; Murthy, N.; Press, O. W.; Tirrell, D. A.; Hoffman, A. S.; Stayton, P. S. *Bioconj. Chem.* **1999**, *10*, 401.

(21) Cho, Y. W.; Kim, J. D.; Park, K. *J. Pharm. Pharmacol.* **2003**, *55*, 721.

(22) Maxfield, F. R. *J. Cell Biol.* **1982**, *95*, 676.

(23) Ogris, M.; Steinlein, P.; Kursa, M.; Mechtler, K.; Kircheis, R.; Wagner, E. *Gene Ther.* **1998**, *5*, 1425.

(24) Zhivotovsky, B.; Orrenius, S.; Brustugun, O. T.; Doskeland, S. O. *Nature* **1998**, *391*, 449.

(25) Cai, J.; Yang, J.; Jones, D. P. *Biochim. et Biophys. Acta* **1998**, *1366*, 139.

TOC Entry

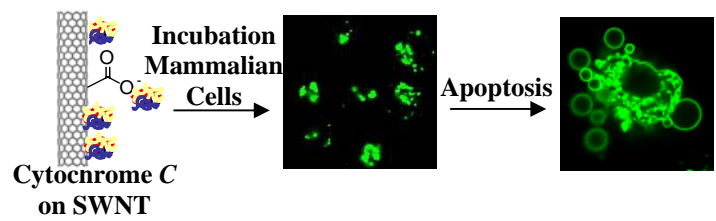